\documentclass[conference,a4paper]{APSIPA2020}\IEEEoverridecommandlockouts
\usepackage{multirow}
\usepackage{graphicx}
\usepackage{amsmath}
\usepackage[psamsfonts]{amssymb}
\usepackage{amsxtra}
\usepackage{threeparttable}
\usepackage{color}
\usepackage{epsfig,amssymb,amsmath}
\usepackage{multirow}
\usepackage{subfigure}
\usepackage{mathrsfs}
\usepackage{float}
\usepackage{hyperref}
\usepackage{url}
\usepackage[ruled,vlined]{algorithm2e}
\usepackage{cite}

\usepackage{CJKutf8}

\begin{document}

\title{Spectrum and Prosody Conversion for Cross-lingual Voice Conversion with CycleGAN}
\author{%
\authorblockN{%
Zongyang Du~\authorrefmark{1},
Kun Zhou \authorrefmark{1},
Berrak Sisman \authorrefmark{2}\authorrefmark{1} and
Haizhou Li \authorrefmark{1}\thanks{\textbf{Speech Samples:} \url{https://kunzhou9646.github.io/cross-vc2020/}}
}
\authorblockA{%
\authorrefmark{1}
Department of Electrical and Computer Engineering, National University of Singapore, Singapore}
\authorblockA{%
\authorrefmark{2}
Information Systems Technology and Design (ISTD) Pillar, Singapore University of Technology and Design, Singapore}
}

\maketitle
\thispagestyle{empty}

\begin{abstract}

Cross-lingual voice conversion aims to change  source speaker’s voice to sound like that of target speaker, when source and target speakers speak different languages. It relies on non-parallel training data from two different languages, hence, is more challenging than mono-lingual voice conversion. Previous studies on cross-lingual voice conversion mainly focus on spectral conversion with a linear transformation for F0 transfer. However, as an important prosodic factor, F0 is inherently hierarchical, thus it is insufficient to just use a linear method for conversion. We propose the use of continuous wavelet transform (CWT) decomposition for F0 modeling.  CWT provides a way to decompose a signal into different temporal scales that explain prosody in different time resolutions. We also propose to train two CycleGAN pipelines for spectrum and prosody mapping respectively. In this way, we eliminate the need for parallel data of any two languages and any alignment techniques. Experimental results show that our proposed \textit{Spectrum-Prosody-CycleGAN} framework outperforms the \textit{Spectrum-CycleGAN} baseline in subjective evaluation. To our best knowledge, this is the first study of prosody in cross-lingual voice conversion. 


\end{abstract}
\section{Introduction}
Voice conversion (VC) aims to convert speaker characteristics from source speaker to target speaker \cite{sisman2020overview}. It is an enabling technique for many applications, such as text-to-speech synthesis \cite{liu2019teacher,liu2020wavetts} and emotional voice conversion \cite{Zhou2020, zhou2020converting}. 

\begin{figure}[ht]
\centering
\subfigure[Training phases of mono-lingual and cross-lingual voice conversion.]{
\begin{minipage}[t]{1\linewidth}
\centering
\includegraphics[width=9cm]{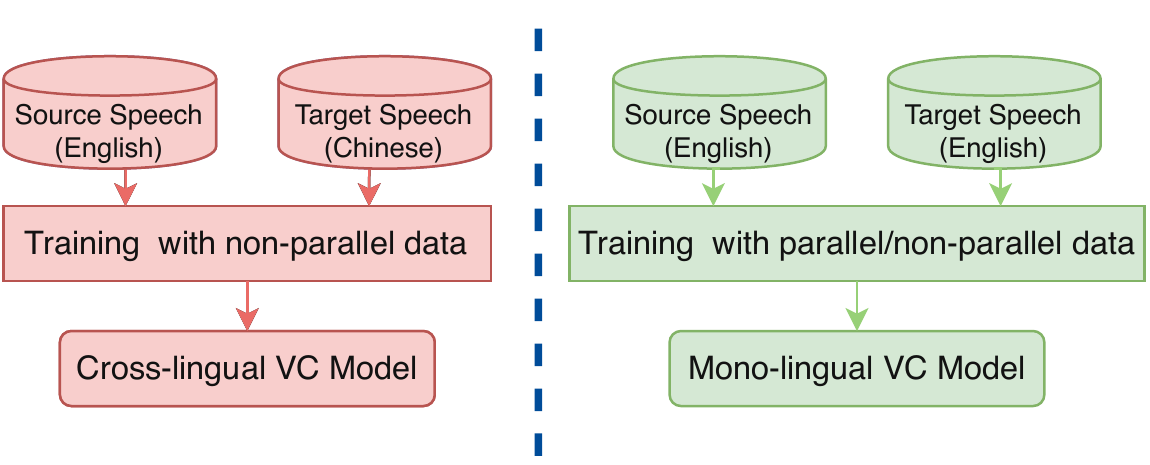}
\label{fig:cross-lingual_training}
\end{minipage}
}

\subfigure[Mono-lingual voice conversion converts source to target speech in the same language at run-time. However, cross-lingual voice conversion converts an English utterance from an English speaker to a Chinese speaker, or a Chinese utterance from a Chinese speaker to an English speaker.
Taking the former as an example, we would like the converted voice to sound like a native English speaker with voice characteristics similar to the Chinese speaker.]{
\begin{minipage}[t]{1\linewidth}
\centering
\includegraphics[width=9cm]{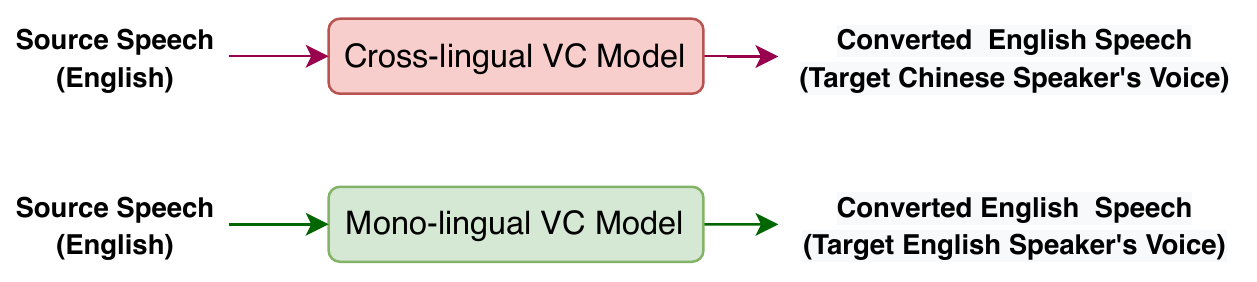}
\label{fig:cross-lingual_conversion}
\end{minipage}%
}%
\centering
\vspace{-1mm}
\caption{A comparison between cross-lingual and mono-lingual voice conversion in the training and conversion phase. Red boxes represent cross-lingual voice conversion, while green boxes represent mono-lingual voice conversion.}
\label{fig:cross-lingual}
\end{figure}

In this paper, we focus on cross-lingual voice conversion~\cite{mohammadi2017overview}, where the source and target speakers speak  different languages. Cross-lingual voice conversion is more challenging  than mono-lingual voice conversion~\cite{sisman2019group} because source and target speakers speak in two different phonetic systems and prosodic styles. Furthermore,  parallel training data is not easily available \cite{zhou2019cross,mashimo2002cross, sisman2019study}. The comparison between cross-lingual and mono-lingual voice conversion is illustrated in Fig. \ref{fig:cross-lingual_training}. and Fig. \ref{fig:cross-lingual_conversion}. In the training phase, cross-lingual voice conversion is trained with non-parallel speech from the source and target speaker speaking with different languages, whereas mono-lingual voice conversion can be trained with parallel/non-parallel speech from source and target speaker speaking with the same language \cite{Kaneko2017ParallelDataFreeVC,mingyang2020deepconversion,Shikano1991SpeakerAA,Abe1988VoiceCT}.

Previous studies on mono-lingual voice conversion mainly focus on finding a mapping function between the source and target through parallel training data, which includes
Gaussian mixture model (GMM) \cite{toda2007voice}, non-negative matrix factorization (NMF) based sparse representation \cite{ccicsman2017sparse, sisman2018voice}, and group sparse representation \cite{sisman2019group}. Recent deep learning approaches, such as deep neural network (DNN) \cite{chen2014voice} and generative adversarial network (GAN)~\cite{sisman2018adaptive}, have greatly improved conversion quality. 


One way of cross-lingual voice conversion is to rely on multilingual speakers to provide same-speaker, cross-lingual training data.
Some statistical approaches for spectral mapping include codebook mapping \cite{abe1991statistical} and Gaussian mixture model (GMM)~\cite{mashimo2002cross}, that show comparable performance with that of mono-lingual voice conversion. However, collecting such data from a bilingual speaker can be expensive and time-consuming. Besides, system performance also depends largely on the speaker's proficiency in the language pair. To circumvent the need for bilingual data, hidden Markov model (HMM) \cite{qian2011frame}, unit selection \cite{sundermann2006text,wang2015aa}  and the iterative frame alignment \cite{erro2007frame,erro2009inca} were proposed to find source-target frame pairs from non-parallel utterances.

More recently, GAN-based methods, including cycle-consistent adversarial network (CycleGAN) \cite{kaneko2017parallel,fang2018high,lorenzo2018can} and variational autoencoding wasserstein generative adversarial networks (VAW-GAN) \cite{hsu2017voice,zhou2020seen} have achieved high-quality mono-lingual voice conversion with non-parallel data. GAN-based methods  for cross-lingual voice conversion~\cite{sisman2019study} have shown similar results with monolingual voice conversion without the need for parallel training data, nor external modules (such as ASR).

Unfortunately, prosody conversion for cross-lingual voice conversion has not been well studied. The differences between two languages lie not only in phonetic systems, but also in linguistic prosody and speaking style, that are characterized by the F0 contour of speech. Motivated by the success of CycleGAN in spectral conversion, we would like to study the conversion of both spectrum and prosody in this paper. In the context of cross-lingual voice conversion, the source speaker is a native speaker of the language, therefore, the source linguistic prosody, such as sentence-level pitch trajectory, should be carried over to the target as much as possible, while the conversion is expected to handle speaker-dependent prosody elements such as pitch level, phoneme-level pitch patterns etc. 

It is well known that speaker characteristics are usually related to: 1) prosodic factors which are concerned with syllables and larger units of speech instead of individual phonetic segments (vowels and consonants); and 2) segmental factors that involve short time spectrum. As an essential prosodic factor, fundamental frequency (F0) represents the variation of the vocal pitch over the whole time domains \cite{ming2015fundamental}. Previous voice conversion studies propose to convert F0 with a simple linear transformation \cite{zhou2019cross} or some statistical methods such as GMM \cite{aihara2012gmm}. Since F0 is hierarchical in nature and affected by short-term as well as long-term dependencies, we believe that these methods are insufficient to model the variations of F0 in different temporal scales~\cite{sisman2018wavelet, Zhou2020}. Therefore, we propose to use CWT decomposition to analyse F0 in different time scales, and find a mapping for each time scale. The CWT decomposition describes a frame-based F0 value with a set of CWT coefficients, that represent  prosodic features. We note that this is the first study of F0 modeling for cross-lingual voice conversion. 


In this paper, we propose a cross-lingual voice conversion framework based on CycleGAN, that can learn the mappings of spectrum and prosody between the speakers speaking with different languages. We also use CWT to decompose F0 into 10 different temporal scales to describe the prosody from the phone level to the utterance level. It is noted that our proposed framework does not rely on any training data from bilingual speakers or any other external modules such as speech recognition or time alignment procedures.

The main contributions of this paper include: 1) we propose a cross-lingual voice conversion framework based on CycleGAN to convert the spectrum and prosody; 2) we propose to analyze F0 in different time resolutions with CWT; and 3) we explored the effect of prosody conversion in cross-lingual voice conversion. To our best knowledge, this paper reports the first attempt to incorporate generative models and CWT-based prosody analysis for cross-lingual voice conversion.

The rest of this paper is organized as follows: In Section II, we provide the motivation and related work to set the stage for this study. In Section III, we propose a spectro-prosodic cross-lingual voice conversion framework. In Section VI, the experiment results are presented. Section V concludes the discussion.

\begin{figure}[htbp]
\flushleft
\subfigure[10 scales of CWT coefficients of an English utterance by one female speaker: "And to wonder what was going to happen next".]{
\begin{minipage}[t]{1\linewidth}
\centering
\includegraphics[width=8.8cm]{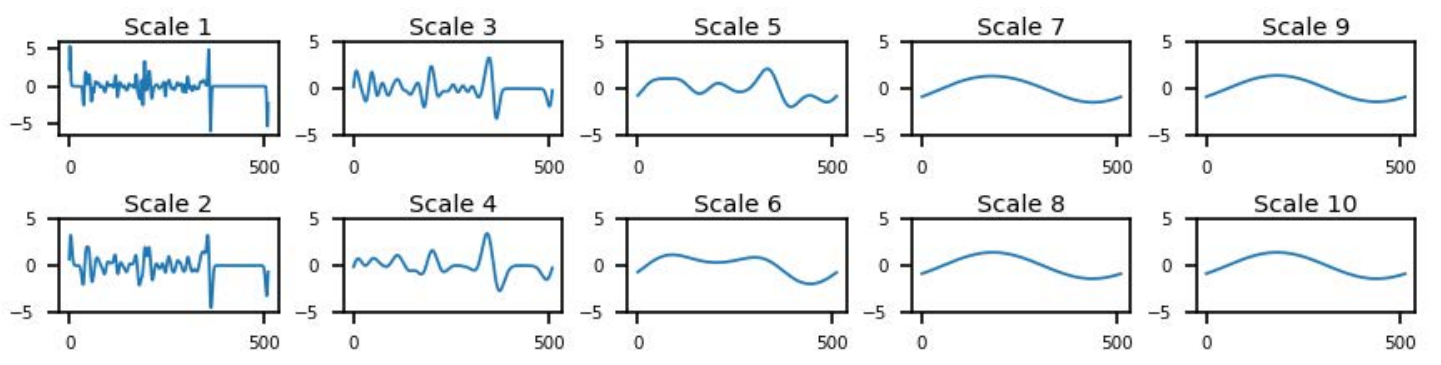}
\label{fig:SF2}
\end{minipage}%
}%

\subfigure[10 scales of CWT coefficients of an English utterance by another female speaker: "And to wonder what was going to happen next".]{
\begin{minipage}[t]{1\linewidth}
\centering
\includegraphics[width=8.8cm]{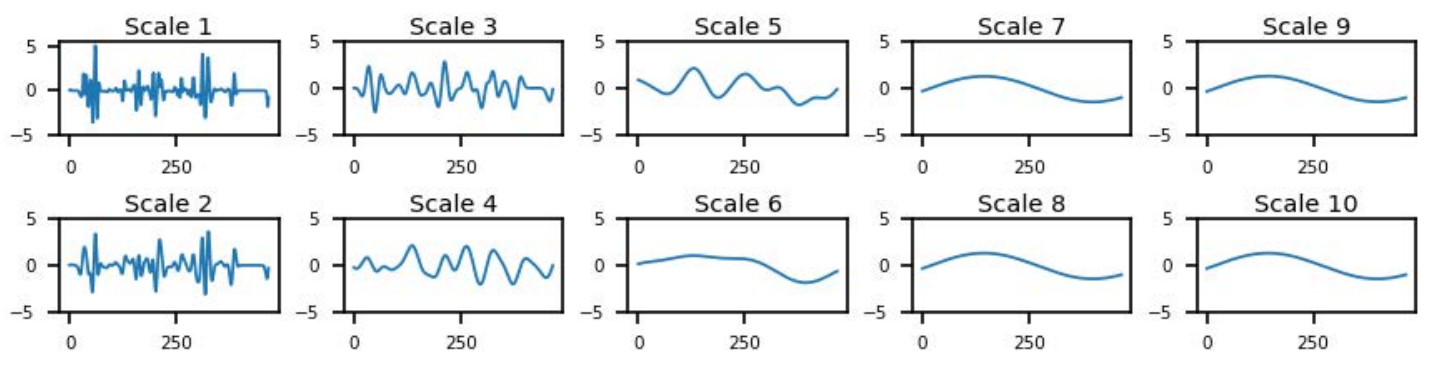}

\label{fig:TF2}
\end{minipage}%
}%

\centering

\caption{Illustration of 10-scale CWT coefficients of the same  utterance but spoken by two different female speakers. Lower scales capture the short-term variations (such as phoneme) and higher scales capture the long-term variations (such as sentence).}

\label{fig:cwt_examples}
\end{figure}

\section{Related Work}

\subsection{Prosody Modelling with CWT}
Generally speaking, speech can be characterized by spectral and prosodic features \cite{helander2007analysis, morley2012synthetic}. Prosody is supra-segmental and hierarchical in nature, hence prosody modelling is not as straghtforward as frame-based spectral modelling in voice conversion.  

Continuous wavelet transform (CWT) has shown to be effective in simultaneous analysis and visualization of various time scales of a signal \cite{vainio2013continuous}. It provides a way to describe F0 in different time scales. We also note that CWT has been successfully applied in analysis and modelling of F0 in speech synthesis\cite{vainio2013continuous,tokuda2013speech,suni2013wavelets} and mono-lingual voice conversion \cite{csicsman2017transformation,sisman2018wavelet}. In this paper, we further this idea and decompose F0 into 10 temporal scales.

\begin{figure}[htbp]
\flushleft
\subfigure[The original F0 contour]{
\begin{minipage}[t]{0.5\linewidth}
\centering
\includegraphics[width=4.2cm]{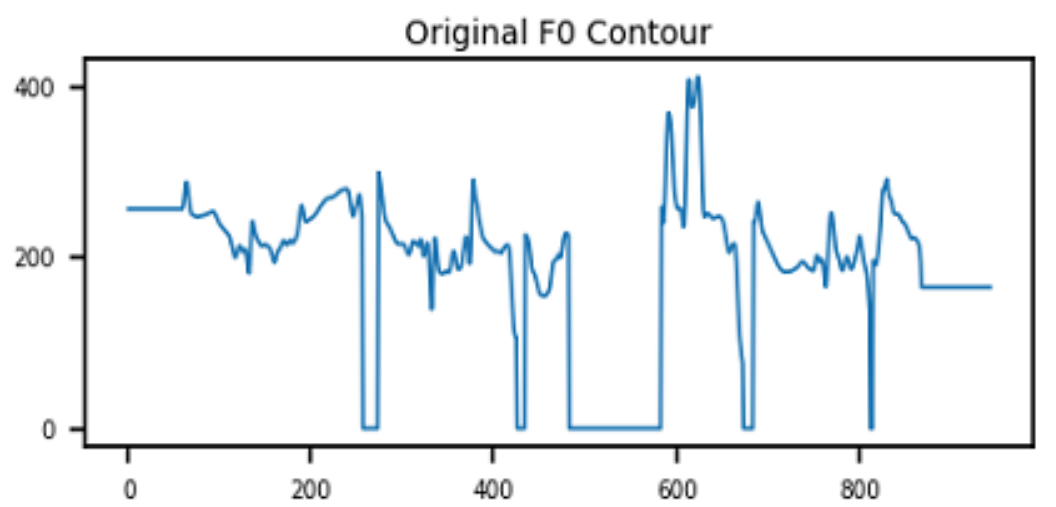}
\label{fig:original F0}
\end{minipage}%
}%
\subfigure[F0 contour after pre-processing ]{
\begin{minipage}[t]{0.5\linewidth}
\centering
\includegraphics[width=4.2cm]{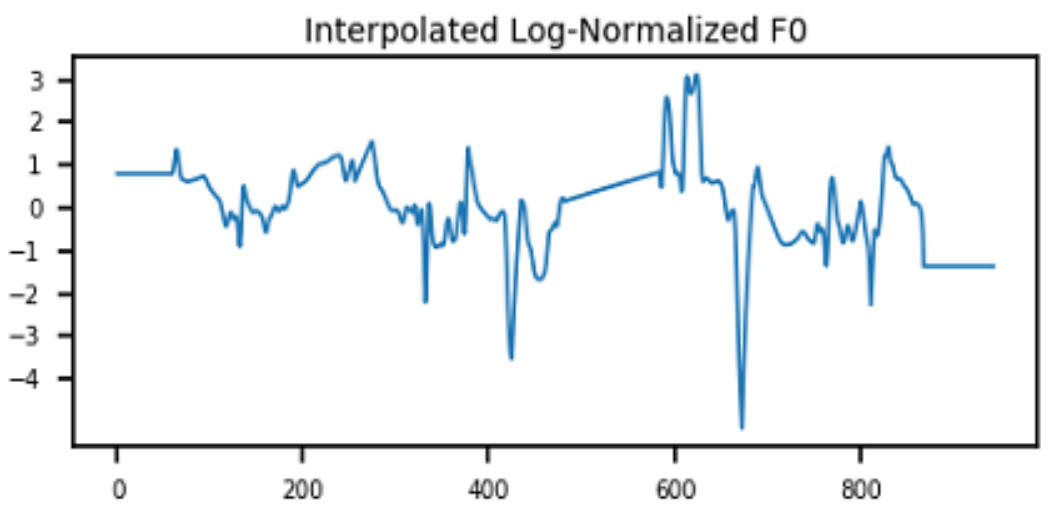}
\label{fig:continuous F0}

\end{minipage}%
}%

\subfigure[ 10-scale CWT coefficients of F0. ]{
\begin{minipage}[t]{1\linewidth}
\centering
\includegraphics[width=8.8cm]{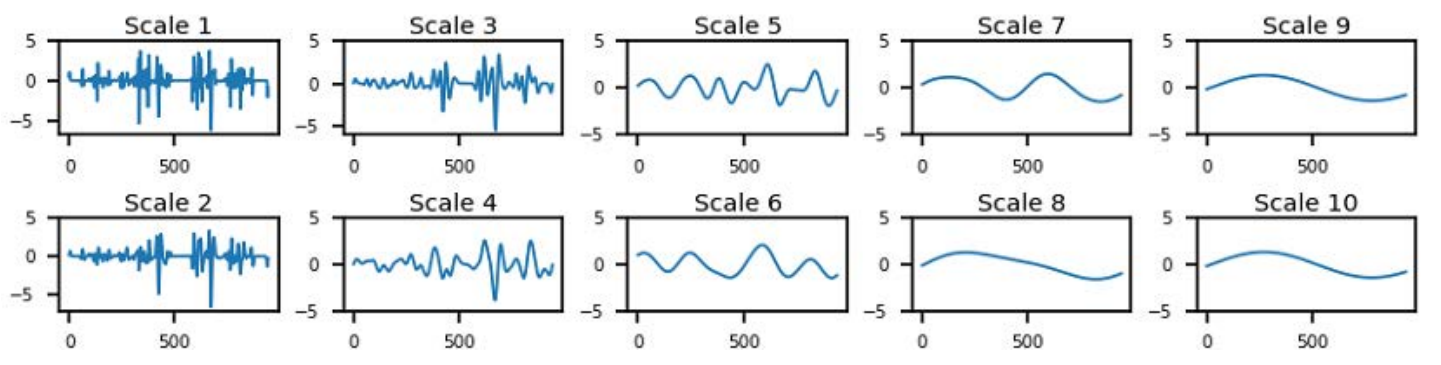}

\label{fig:cwt_english}
\end{minipage}%
}%

\centering
\caption{ 10-scale CWT coefficients of an English utterance spoken by the source speaker: ``Thousands of men were rushing into the Northland".}

\label{fig:cwt_10}
\end{figure}

Given a continuous input signal $f$, its continuous wavelet transform $ W(\tau,t)$ can be written as:
\begin{equation}
    W(\tau,t)=\tau^{-1/2}\int_{-\infty}^{+\infty}f(x)\psi(\frac{x-t}{\tau})dx
\end{equation}
where $\psi$ is the Mexican hat mother wavelet. The original signal $f$ can be recovered from the wavelet representation $W(\tau,t)$ by inverse transform using the following formula:
\begin{equation}
f(t)=\ \int_{-\infty}^{+\infty}\int_{0}^{+\infty}W\left(\tau,t\right)\tau^{-5/2}\psi\left(\frac{x-t}{\tau}\right)dxd\tau
\end{equation}
Suppose that we decompose the input signal $f$ into 10 scales~\cite{ming2016exemplar, ming2016deep}, $f$ can be represented by 10 separate components given by:
\begin{equation}
W_i(f,t)=\\ W(f)(2^{i+1}\tau_0,t)(i+2.5)^{-5/2}
\end{equation}
where $i=1,...,10$ and $\tau_0 = 5 ms$. These time scales were originally proposed in \cite{suni2013wavelets}.  

Given 10 wavelet components, $\hat{W}_i(f,t)$, that are the converted version of CWT components for target speaker, we can recompose signal $\hat{f}$  by the following formula \cite{ming2016deep}:

\begin{equation}
    \hat{f}(t) = \sum_{i=1}^{10}\hat{W}_i(f, t)(i+2.5)^{-5/2}
\label{eq:restruct}
\end{equation}

As can be observed in Fig. \ref{fig:SF2} and \ref{fig:TF2} where two speakers read the same text, high scale (e.g. scale 9 and 10) coefficients  are similar between speakers, while low scale (e.g. scale 1 and 2) coefficients are speaker-dependent~ \cite{sisman2018wavelet}.  



\begin{figure}[t]
    \centering
    \includegraphics[width=8.8cm]{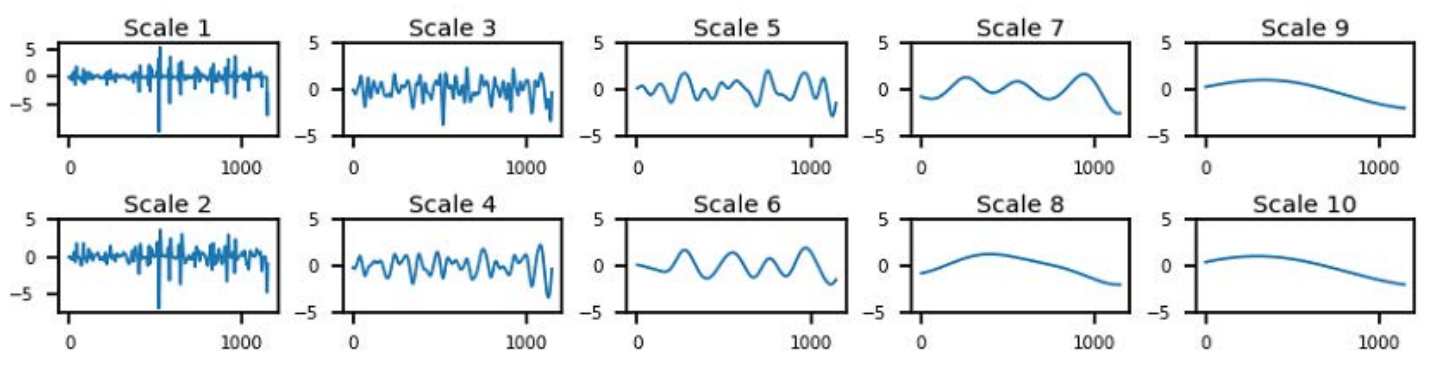}
    \vspace{-2mm}
    \caption{10-scale CWT coefficients of a Chinese utterance spoken by the target speaker: \protect\begin{CJK}{UTF8}{gbsn}“法国人民深深铭记着将军对法兰西民族的丰功伟绩”\protect\end{CJK}.}
    \vspace{-3mm}
    \label{fig:cwt_10_chinese}
\end{figure}

With the multi-scale CWT decomposition, we can now represent a F0 sequence  using a sequence of CWT coefficient frames, in parallel to spectral frames of an utterance. We expect to train a prosody mapping function to learn the mapping between individual CWT coefficients, in particular to reflect, the speaker style transfer.  Fig. \ref{fig:cwt_english} and \ref{fig:cwt_10_chinese} show an example of English-Chinese training pair from two different speakers. We will study the prosody mapping in  Section III.


\subsection{CycleGAN for Style Transfer}
Generative adversarial networks (GANs) \cite{goodfellow2014generative,  ak2019deep, ak2019attribute} provide a way of representing and modelling the high-dimensional distribution of data. GANs  learn deep representation and generate many different acceptable answers by implicitly modelling the high-dimensional distribution of data \cite{emir2019semantically}. GANs consist of two neural networks, a generator and a discriminator, which compete with each other. The generator creates samples following the same distribution of the training set to fool the discriminator while the discriminator distinguishes whether they are real samples from training set or fake generated samples \cite{kameoka2018stargan, ak2020incorporating}. GANs have shown remarkable performance in the fields of computer version \cite{isola2017image, zou2020edge, zhu2017unpaired,chen2017show, aklearning, ak2018learning}, natural language processing \cite{li2017adversarial,yang2017improving}, speech synthesis \cite{guo2019new, lorenzo2018can, zhao2018wasserstein} and voice conversion \cite{fang2018high, sisman2019study, sisman2018adaptive}.

CycleGAN, a successful implementation of GANs, was first proposed for unpaired image-to-image translation \cite{zhu2017unpaired}, and then applied to other non-parallel style transfer tasks, such as speech synthesis and mono-lingual voice conversion \cite{kaneko2017parallel,kaneko2018cyclegan,kaneko2019cyclegan}. CycleGAN has also been used for spectrum mapping in cross-lingual voice conversion \cite{sisman2019study}.  In this paper, we extend the idea and formulate a CycleGAN that learns prosody mapping of CWT coefficients without the need of parallel or bilingual data. Despite different languages, as shown in Fig. 3(c) and Fig. 4, the low scale CWT coefficients of F0 between the two utterances could be very different, and the high scale CWT coefficients of F0 are similar. We expect that CycleGAN is able to learn the differentiated mapping relationship between low and high scale coefficients.

\section{CycleGAN based cross-lingual voice conversion}


 In this section, we propose a CycleGAN-based cross-lingual voice conversion framework that effectively learns a mapping between source and target speakers from two different languages. We also use CWT to decompose F0 into 10 different time scales for both source and target speakers, ranging from the micro-prosody level to the whole utterance level, with the aim to describe the prosody in different time resolutions. 

\begin{figure*}

    \centering
    \includegraphics[scale = 0.8]{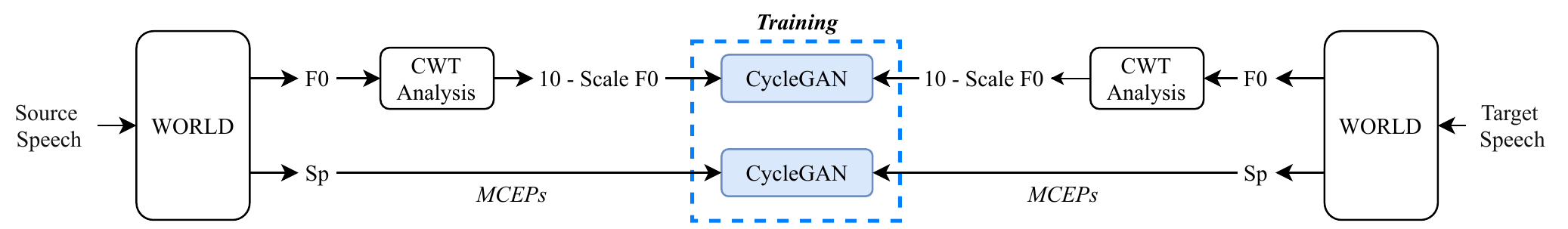}
 
    \caption{The training phase of the proposed CycleGAN-based cross-lingual VC framework. Blue box represents the cross-lingual VC model for prosody conversion called \textsl{F0-CycleGAN} and yellow box is the cross-lingual VC model for spectrum conversion called \textsl{MCEP-CycleGAN}.}
    \label{fig:training}
\end{figure*}
\begin{figure*}
    \centering
    \includegraphics[scale = 0.8]{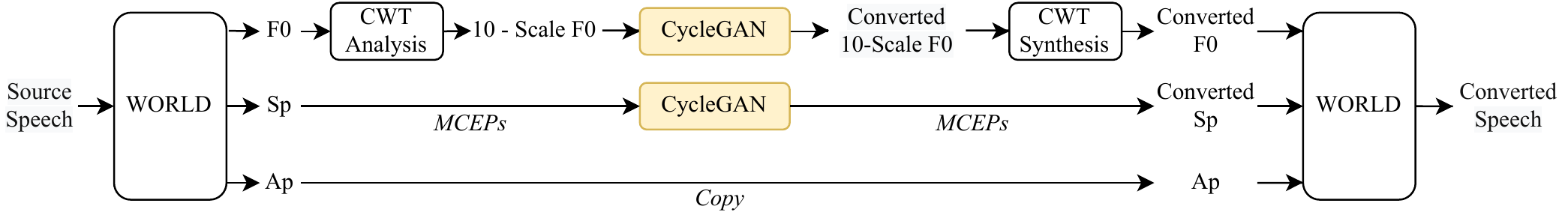}
  
    \caption{The run-time conversion phase of the proposed \textit{Spectrum-Prosody-CycleGAN} framework. Colored boxes  represent the trained models in Fig. \ref{fig:training}.}
    \label{fig:conversion}
\end{figure*}

\subsection{Training of Spectrum and Prosody Conversion with CycleGANs}
We first use WORLD vocoder to extract spectral features and fundamental frequency (F0) from source and target waveform. It is noted that the extracted F0 features are discontinuous due to the voiced/unvoiced activities in an utterance, as shown in Fig. \ref{fig:original F0}. Since CWT is sensitive to the gaps in the extracted F0 contour \cite{ming2015fundamental, csicsman2017transformation}, the following pre-processing steps are necessary \cite{sisman2018wavelet, sisman2019group}: 1) linear interpolation over unvoiced regions; 2) transformation of F0 from linear to a logarithmic scale; and 3) normalizing the resulting F0 to zero mean and unit variance. The continuous F0 signals after pre-processing is shown in Fig. \ref{fig:continuous F0}. After the pre-processing, we obtain the 10-scale CWT-F0 coefficients, and a 10-dimensional prosody frame.

For spectrum conversion, we extract 24-dimensional Mel-cepstral coefficients (MCEPs) features for spectral CycleGAN model, denoted as \textit{Spectrum-CycleGAN}. As for prosody conversion, we use  10-dimensional CWT-F0 coefficients as  prosodic features for  prosodic CycleGAN model, and denoted as \textit{Prosody-CycleGAN}. The same training procedure applies to the training of both models. 

During training, CycleGAN is capable of learning the forward and inverse mappings simultaneously through the non-parallel training data and trying to find an optimal pseudo pair for the spectrum and prosody conversion. The training phase of the proposed framework is shown in Fig. \ref{fig:training}, that includes a spectrum and a prosody modeling pipeline. Previous study on emotional voice conversion \cite{Zhou2020} has shown that separate training of the spectrum and prosody can achieve better performance than  joint training. Therefore, we train the two CycleGANs   separately. 

We assume that source and  target speaker speak two different languages: English and Chinese respectively, which are denoted as $en$ and $cn$. The goal of CycleGAN is to learn a mapping between the source $x_{en}\in X_{en}$ and the target $y_{cn}\in Y_{cn}$. CycleGAN is incorporated with three main loss functions: adversarial loss, cycle-consistency loss, and identity-mapping loss~\cite{kaneko2017parallel}.

An adversarial loss measures how distinguishable between the data distribution of converted data and source or target data. For the forward mapping, it is defined as:
\begin{align}
  &L_{ADV}(G_{X_{en}\rightarrow Y_{cn}},D_{Y_{cn}}) = \nonumber \\ 
   &\mathbb{E}_{y_{cn}} P_{Y_{cn}(y_{cn})}[\log(D_{Y_{cn}}(y_{cn}))] \nonumber \\ 
    +&\mathbb{E}_{x_{en} P_{X_{en}}(x_{en})}\left[\log(1-D_{Y_{cn}}\left(G_{X_{en}\rightarrow Y_{cn}}(x_{en})\right)\right]
\end{align}

Since the adversarial loss only restricts $G_{X\rightarrow Y}(x)$ to follow the target distribution, a cycle-consistency loss is proposed in order to guarantee the consistency of the contextual information between input and output. It is defined as:
\begin{multline}
    L_{CYC}(G_{X_{en}\rightarrow Y_{cn}},G_{Y_{cn}\rightarrow X_{en}}) 
    =\\ \mathbb{E}_{x_{en}\sim P(x_{en})}[\Vert G_{Y_{cn}\rightarrow X_{en}}(G_{X_{en}\rightarrow Y_{cn}}(x_{en}))-x_{en}\Vert_1]\\
    +\mathbb{E}_{y_{cn}\sim P(y_{cn})}[\Vert G_{X_{en}\rightarrow Y_{cn}}(G_{Y_{cn}\rightarrow X_{en}}(y_{cn}))-y_{cn}\Vert_1]
\end{multline}
Cycle-consistency loss encourages the forward mapping $G_{X_{en}\rightarrow Y_{cn}}$ and the inverse mapping $G_{Y_{cn}\rightarrow X_{en}}$ to find an optimal pseudo pair of $(x_{en},y_{cn})$ through circular conversion.

In order to preserve linguistic information without any external processes, identity mapping loss is defined as: 
\begin{align}
    &L_{ID}(G_{X_{en}\rightarrow Y_{cn}},G_{Y_{cn}\rightarrow X_{en}})= \nonumber \\ 
    &\mathbb{E}_{x_{en}\sim P(x_{en})}[\Vert G_{Y_{cn}\rightarrow X_{en}}(x_{en})-x_{en}\Vert]\nonumber \\ 
    + &\mathbb{E}_{y_{cn}\sim P(y_{cn})}[\Vert G_{X_{en}\rightarrow Y_{cn}}(y_{cn})-y_{cn}\Vert]
\end{align}
With these three loss, we expect CycleGAN to learn a bi-directional mapping between the spectrum and prosody distributions in different languages, from source and target speakers.

\subsection{Run-time Conversion}

The conversion phase of the proposed framework is illustrated in Fig. \ref{fig:conversion}. Similar to that of training phase, WORLD vocoder is used to extract spectral features, F0 and aperiodicities (APs) of source speech. We then encode 24-dimensional MCEPs spectral features, and decompose F0 into 10-scales, denoted as CWT-F0 coefficients. The 24-dimensional MCEPs and 10-dimensional CWT-F0 coefficients are converted by the trained \textit{Spectrum-CycleGAN} and \textit{Prosody-CycleGAN} respectively. We reconstruct F0 from the converted 10-dimensional CWT-F0 coefficients through CWT synthesis given in Fig. \ref{fig:conversion}, using Eq. \ref{eq:restruct}. 

At run-time, we present the combined results of spectrum conversion and prosody conversion to the vocoder for reconstruction of speech waveform. Therefore, we call the proposed framework \textit{Spectrum-Prosody-CycleGAN}. 
 It is noted that APs are directly copied from source to that of target speaker without any modification \cite{kaneko2017generative}.

\section{Experiments}

In this section, we conduct experiments to assess the performance of our proposed CycleGAN-based cross-lingual voice conversion framework for spectrum and prosody. We use VCC 2016 dataset \cite{toda2016voice} and  Blizzard Challenge 2010 database \cite{kinga2009blizzard}, which consists of English and Chinese speech data respectively. We choose four female English speakers (SF3, TF2, TF1, SF1), and one female Chinese speaker for experiments.
We conduct cross-lingual voice conversion both from English to Chinese, denoted as \textit{$en2cn$}; and from Chinese to English, denoted as \textit{$cn2en$}, respectively. For \textit{$en2cn$}, we conduct two experiments and choose SF3 and TF2 as the source speakers. For \textit{$cn2en$}, we conduct two experiments and choose SF1 and TF1 as the source speakers, respectively.

We note that bilingual data are required for MCD calculation in cross-lingual VC \cite{sisman2019study, zhou2019cross}, hence we only conduct subjective experiments to show the effectiveness of our proposed framework. 

We build two systems for a comparative study: 1) proposed \textit{Spectrum-Prosody-CycleGAN}, and 2) baseline \textit{Spectrum-CycleGAN} ~\cite{sisman2019study}, where spectrum is converted with CycleGAN and fundamental frequency (F0) is converted through the logarithm Gaussian (LG) normalized transformation \cite{liu2007high}. In \textit{Spectrum-Prosody-CycleGAN}, we perform CWT to decompose F0 into 10 different time scales and train two CycleGAN networks using non-parallel data with two different languages to learn the mappings of spectral and prosody features between source and target speaker.

\subsection{Experimental Setup}
To evaluate the systems under the condition of non-parallel and limited amount of data, we use 81 non-parallel utterances from source and target speakers with English and Chinese for training, and 54 utterances for evaluation. 
It is noted that our proposed cross-lingual VC method is trained under the disadvantageous condition (non-parallel and limited amount of data).

The speech data is downsampled to 16kHz, and 24-dimensional Mel-cepstral coefficients (MCEPs), fundamental frequency (F0), and aperiodicity (APs) are then extracted every 5 ms using WORLD vocoder \cite{morise2016world}. For both frameworks, we extract 24-dimensional MCEPs and one-dimensional F0 features for each frame. We further obtain 10-dimensional CWT-based F0 features from one-dimensional F0 features with CWT analysis in our proposed framework. It is noted that APs are directly copied from the source to target without any modification.

In both systems, the generator uses a one-dimensional (1D) CNN to capture the relationship among the overall features while preserving the temporal structure. The 1D CNN is incorporated with down-sampling, residual,and up-sampling layers. We design the discriminator using a 2D CNN to focus on a 2D spectral texture. In the training phase, we set $\lambda_{CYC}$=10. and only use $L_{ID}$ for the first $10^{4}$ iterations.  The Adam optimizer \cite{kingma2014adam} with a batch size of 1 is used to train the networks. The initial learning rate for the generators is set to 0.0002 while that of discriminators are set to 0.0001. The learning rate keeps the same for the first $2 \times 10^{5}$ iterations, and  linearly decays over the second $2 \times 10^{5}$ iterations. The momentum term $\beta_1$ is set to 0.5.

\subsection{Evaluations}
We conduct two listening tests to assess system performance of  in terms of voice quality and speaker similarity. For each test, we conduct cross-lingual VC from English to Chinese and from Chinese to English respectively, which denoted as  \textit{$en2cn$} and \textit{$cn2en$}.
15 bilingual native Chinese speakers participated in all the listening tests and listened to 120 converted utterances in total.

\subsubsection{Mean Opinion Score Tests}
To evaluate the converted speech quality, we first conduct mean opinion score (MOS) test between the baseline and our proposed method. 15 sentences are randomly selected from the evaluation set. In MOS test, each subject are asked to score each sample in a five-point-scale (5: excellent, 4: good, 3: fair, 2: poor, 1: bad). As shown in Table \ref{tab:table1}, our proposed method achieves comparable results with the baseline 
for both  \textit{$en2cn$} and \textit{$cn2en$} . 

\begin{table}[t]
    \centering
    \caption{A comparison of the MOS results between baseline and our proposed method for English-to-Chinese (en2cn) and Chinese-to-English (cn2en). }
    \begin{tabular}{|c||c|c|c|c}
    \hline
    \multirow{2}{*}{Framework} & \multicolumn{2}{c|}{Mean	Opinion	Score	(MOS)}  \\ \cline{2-3} 
                  & \textit{$en2cn$} & \textit{$cn2en$} \\ \hline
    Spectrum-CycleGAN \& LG \cite{sisman2019study} & 2.91 $\pm$ 0.33  & 3.30 $\pm$ 0.27              \\ \hline
    \textbf{Spectrum-Prosody-CycleGAN} & \textbf{2.92 $\pm$ 0.27}  & \textbf{3.34 $\pm$ 0.29 }            \\ \hline

    \end{tabular}
    \label{tab:table1}
\end{table}

\subsubsection{Similarity Listening Tests}

We consider that how similar the converted speech is to the target speech reflects the effect of prosody conversion. Therefore, we conduct XAB preference test between our proposed framework and the baseline framework in terms of speaker similarity. The subjects are asked to listen to the reference target samples and the converted samples of the baseline and our proposed framework, and choose the one which sounds closer to the target sample. We expect that the proposed prosody conversion approach will remarkably improve the speaker similarity, as prosody contains the information of speaking style.


\begin{figure}[t]
    \centering
    \includegraphics[width=8.5cm]{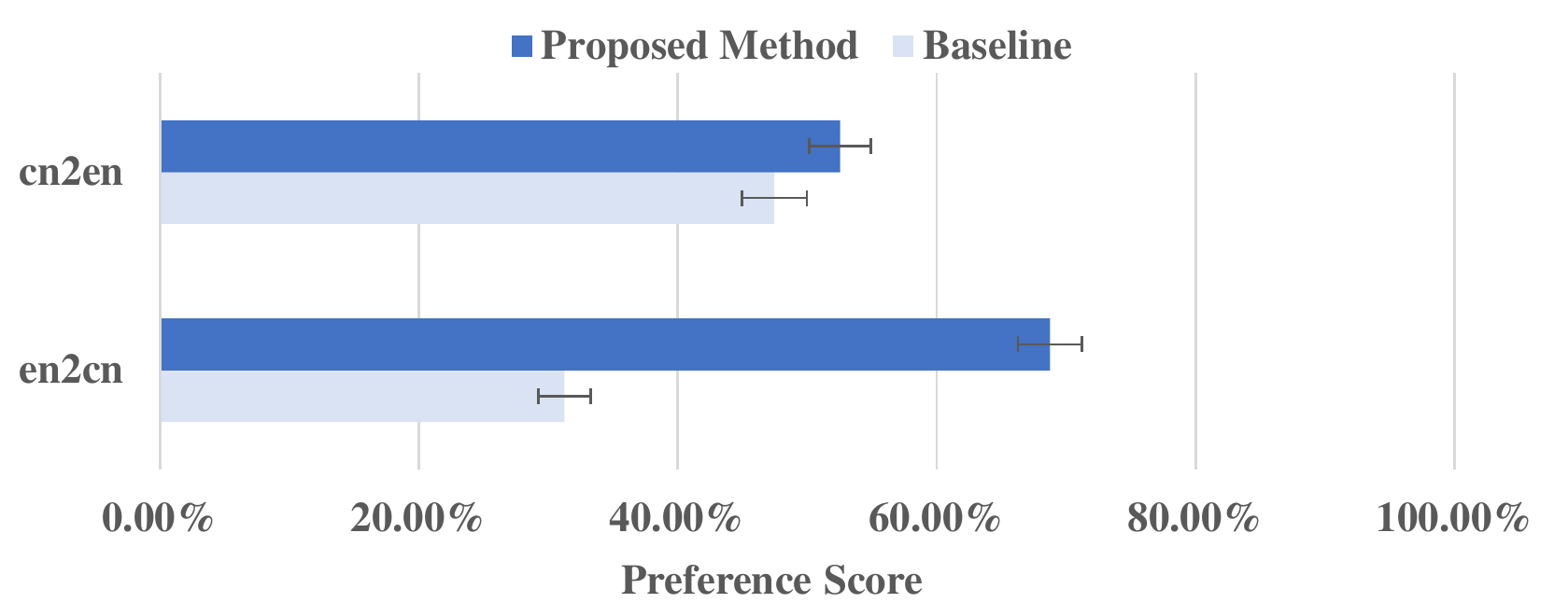}
    
    \caption{The XAB preference results with 95\% confidence interval between the baseline and the proposed cross-lingual framework for speaker similarity. }
    \label{fig:pref1}
    
\end{figure}
As shown in Fig. \ref{fig:pref1}, 
we observe that our proposed method clearly outperforms the baseline framework in both experiments English to Chinese (\textit{$en2cn$}), and Chinese to English (\textit{$cn2en$}) experiments. The results suggest that CWT is an effective way of prosody modeling for cross-lingual voice conversion, and through CycleGAN we can learn a prosody mapping between source and target that speak different languages. The results also validate our idea to use CycleGAN to learn the differentiated mapping relationship among low and high scale coefficients of F0.




\section{Conclusion}
In this paper, we propose a novel parallel-data-free cross-lingual voice conversion framework. We convert the spectrum and prosody based on CycleGAN with non-parallel and limited amount of training data. Moreover, We also provide a non-linear method which uses CWT to describe the prosody in different time scales for cross-lingual voice conversion. Experimental results show the effectiveness of our proposed framework in terms of voice quality and speaker similarity.


\section{Acknowledgement}


This work is supported by the National Research Foundation, Singapore under its AI Singapore Programme (AISG Award No: AISG-100E-2018-006, AISG-GC-2019-002),  the National Robotics Programme (Programmatic Grant Number: 192 25 00054), Human Robot Collaborative AI for AME Programmatic Grant (Programmatic Grant No. A18A2b0046) and Neuromorphic Computing Programmatic Grant (Programmatic Grant No. A1687b0033) from the Singapore Government’s Research, Innovation and Enterprise 2020 plan in the Advanced Manufacturing and Engineering domain. Any opinions, findings and conclusions or recommendations expressed in this material are those of the author(s) and do not reflect the views of National Research Foundation, Singapore.

This work is also supported by SUTD Start-up Grant Artificial Intelligence for Human Voice  Conversion  (SRG  ISTD  2020  158)  and  SUTD  AI  Grant  titled 'The Understanding and Synthesis of Expressive Speech by AI' (PIE-SGP-AI-2020-02).

\bibliographystyle{IEEEtran}
{\footnotesize
\bibliography{apsipa_Zongyang}}

\end{document}